# The effect of RF-DC plasma N$_2$-H$_2$ in the selective hardening process for micro-patterned AISI420

Hengky Herdianto, Dionisius Johanes Djoko Herry Santjojo, Masruroh

# The effect of RF-DC plasma N$_2$-H$_2$ in the selective hardening process for micro-patterned AISI420

Hengky Herdianto, D. J. Djoko H. Santjojo, and Masruroh

View Online    Export Citation

# The Effect of RF-DC Plasma N$_2$-H$_2$ in the Selective Hardening Process for Micro-Patterned AISI420

Hengky Herdianto[a)], D.J. Djoko H. Santjojo, Masruroh

*Department of Physics, Faculty of Mathematics and Natural Science, University of Brawijaya*
*Jl. Veteran, Malang 65145, Jawa Timur, Indonesia*

[a)]Corresponding author: hengkyherdianto@gmail.com

**Abstract.** The high density of RF-DC plasma N$_2$-H$_2$ was used to make precise micro-texturing onto AISI420 has complex textured geometry. The original 2D micro-patterns were drawn onto substrate surface by maskless patterning using of nano-carbon ink. These micro-patterned specimens were further plasma-nitrided at 673 K for 5.4 ks by 70 Pa using the hollow cathode device. The emissive light spectroscopy shows species in plasma were nitrogen atoms together with NH radicals and nitrogen molecular ions. Unprinted surface areas had selectively nitrided, have high nitrogen solute contents up to 12 mass%. Masked area just corresponded to carbon-mapping from printed nano-carbon inks, while unprinted surface to nitrogen mapping. The hardness profile had stepwise change across the borders between these printed and unprinted areas; e.g., the hardness on unprinted surface was 1200 Hv while it remained to be 350 Hv on printed surface. This selective nitriding and hardening enabled to construct the 3D textured miniature dies and products by chemical etching of printed area. These two peaks were related to extended martensitic lattice by high nitrogen extraordinary solid solution. The phase transformation from martensitic lattice $\alpha$'-Fe through expanded phase into $\varepsilon$-Fe$_3$N lattice.

## INTRODUCTION

The case-hardened AISI420 martensitic stainless steels, graded by STAVAX in commercial is used as a mold for injection molding [1]. Micro- and nano-textures on the metal and polymer surfaces of parts and components have functions to reduce the friction and wear in [2] and to improve the joining strength in [3]. As needed, the surface texturing is also employed to have the surface profile leather-touched or textile-patterned. In order to fabricate these micro- and nano-textured surface of products, the mold for precise injection molding or the dies for fine stamping must have their original micro- and nano-textures on their surfaces.

In [1,4-6], the non-traditional processing was proposed to make micro-texturing into dies and molds with aid of the high density RF-DC Plasma N$_2$-H$_2$. The original two dimensional micro-patterns are directly drawn onto the mold and die surfaces. This printed mold and dies are subjected to the plasma nitriding by using the printed area as a mask. Since the masked area is free from the plasma nitriding, the unmasked or unprinted area is selectively plasma nitrided and hardened. The masked or printed area, which has much lower hardness than the nitrided area, is mechanically removed by the blasting or brushing method to finish the micro-textured mold and die surfaces.

Different from the micro-milling and micro-EDM, no CAM data or no tools are necessary to make micro-texturing. Tedious and time-consuming milling and electric discharging processes are saved to fabricate the molds for injection molding in much shorter processing time. In those previous works, the size of micro-pattern in width was limited by 50 to 100 μm; the possibility of fine micro-texturing by the plasma nitriding must be demonstrated with use of the well-aligned micro-patterns.

In the present paper, the maskless patterning is utilized to draw the initial micro-patterns onto the stainless steel specimen and molds. These substrates are plasma nitrided at 673 K to demonstrate their unprinted surfaces are selectively nitrided and hardened. The contour of the nitrogen mapping just corresponds to the initial micro-pattern; the printed micro-patterns on the substrate are textured into the stainless steels as a high content nitrogen solute map.

Hydrogen addition plays important role and beneficial effects in plasma $N_2$-$H_2$ process due to its ionization, dissociation and de-oxidation abilities [7].

# EXPERIMENTAL

## Maskless Patterning with Use of Nano-Carbon Ink

The maskless micro-patterning method was utilized to draw the initial two dimensional micro-patterns directly onto the stainless steel substrate surface. Figure 1 depicts a typical micro-pattern drawn on the stainless steel sheet with use of the nano-carbon ink. Even when using the same CAD data for micro-patterning, there are two manners in drawing: the positive-image drawing as depicted in Figure 1a) and the negative-image drawing as shown in Figure 1b).

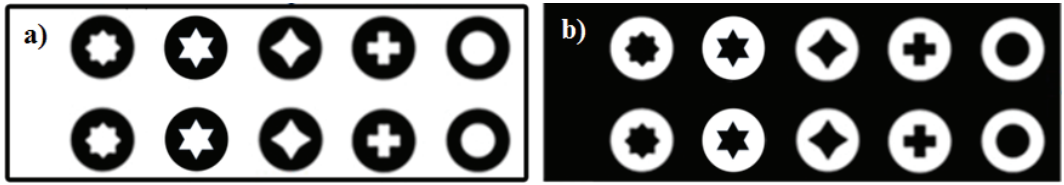

**FIGURE 1**. Micro-patterning onto the stainless sheet surface: a) positive image, b) negative image

At present, the drawing area is limited by 25 x 25 $mm^2$ with the spatial resolution down to 1 μm. The various kinds of micro-units with geometric complexity are straight forwardly designed by CAD and drawn on the AISI420 substrates. This maskless micro-patterning is useful to draw any designed micro-head pattern with the complex geometries and within the high spatial resolution.

## RF-DC Plasma $N_2$-$H_2$ System

RF-DC plasma $N_2$-$H_2$ system was set-up for selective nitriding and hardening of stainless steels. Different from the PECVD (Plasma Enhanced Chemical Vapor Deposition) or DC-pulse nitriding, where the plasmas are ignited and generated in the frequency of 13.56 MHz or its multiples, the present RF-DC plasma $N_2$-$H_2$ system has no mechanical matching box with slow response time of 1 s to 10 s to adjust the applied power. Since both the input and output powers are automatically matched by frequency adjustment around 2 MHz, the matching response time is only limited to 1 ms at most. This prompt power control provides to make full use of mesoscopic plasma pressure range over 50 Pa.

The differences from the conventional processes, the vacuum chamber is electrically neutral so that RF-power and DC-bias should be controlled independently from each other. A dipole electrode is utilized to generate RF-plasma; DC bias is directly applied to the specimens. Heating unit is located under this DC-biased cathode plate. In the following nitriding experiments, the specimens are located on the cathode table before evacuation down to the base pressure of 0.1 Pa.

**TABLE 1**. Summarizes the experimental conditions for present plasma $N_2$-$H_2$

| Process | Parameters |
|---|---|
| Pre-sputtering | DC (450 V), RF (0 V) |
| | Pressure (70 Pa) |
| | Temperature (Room standart) |
| | Duration (400 s) |
| | Carrier gas ($N_2$ only) |
| Nitriding | DC (300 V), RF (250 V) |
| | Pressure (70 Pa) |
| | Temperature (673 K) |
| | Duration (5.4 ks) |
| | Carrier gas ($N_2$ + $H_2$) with partial pressure ($N_2$ = 100 ml/min; $H_2$=20 ml/min) |

Then, nitrogen gas is first introduced as a carrier gas for heating. After heating to the specified holding temperature, the nitrogen pre-sputtering is started at the constant pressure. After pre-sputtering, the hydrogen gas is added to nitrogen gas with the specified partial pressure ratio. In particular, the partial pressure ratio of nitrogen to hydrogen gases is constant by 5 to 1 as a standard condition by controlling the gas flow rate to be 100 ml/min for $N_2$ gas and 20 ml/min for $H_2$ gas, respectively the holding temperature is 673K detailed are presented in Table 1. Both of pressure ($P$) and temperature ($T$) controls are automatically performed with the tolerance of $\Delta P < 1$ Pa in deviation of partial pressure and $\Delta T < 1$ K in temperature fluctuation.

## Observation and Measurement

The plasma condition was observed and identified by using the optical emissive-light spectroscopy (EOS; PMA-11, Hamamatsu, Co. Ltd). SEM (Scanning electron microscope; Shimadzu, Corp) was used to observe printed mask pattern of specimen before nitriding. EDX (Energy Dispersive X-ray; SSX-550, Shimadzu, Co. Ltd.) was utilized to analyze the nitrogen content distribution at the vicinity of the border between the printed and unprinted surface. The hardness of nitride specimen was measured by Vickers micro-hardness testing (Mitsutoyo, Co. Ltd) while XRD (Smartlab, Rigaku, Co., Ltd.) was used for analyzing the microstructure of specimen.

## RESULTS AND DISCUSSION

### Species of Plasma $N_2$-$H_2$

Through the quantitative plasma diagnosis, the generated species in the plasmas are detected and monitored by EOS. Figure 2 and table 2 shows an emissive light spectrum measured in the plasmas for nitriding with and without use of the hydrogen gas. NH radicals as well as the activated nitrogen atom and molecules are detected in both spectra. With use of the mixed nitrogen and hydrogen gases, every peak intensity is enhanced even with a little addition of hydrogen gas. This significant population of NH radicals and $N_2^*$ might be responsible for high infiltration of nitrogen atoms into the stainless steels.

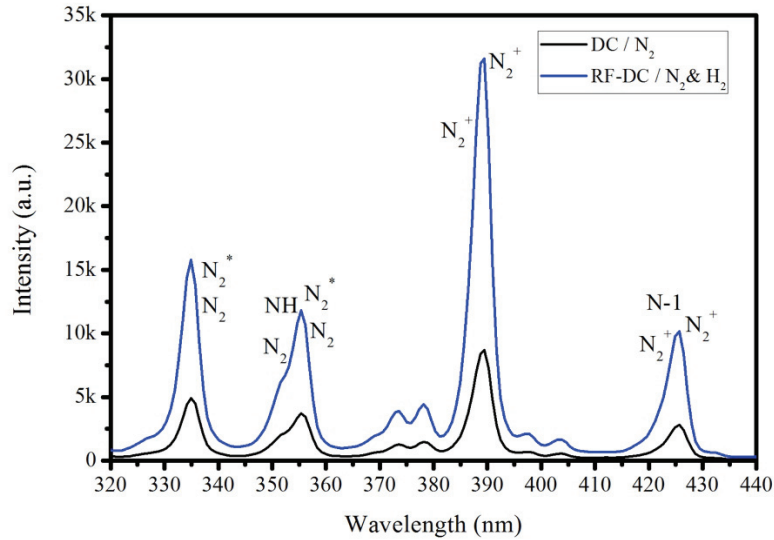

**FIGURE 2**. The emissive light spectra of plasmas with and without addition of hydrogen gas

**TABLE 2.** Species of plasma $N_2$-$H_2$

| Species | Transition | $\lambda$ (nm) | E (eV) | Reference |
|---|---|---|---|---|
| $N_2^+$ | $B^2\Sigma_u^+ \rightarrow X^2\Pi_g^+ \ (1\rightarrow1)$ | 388.43 | | [8] |
| | $B^2\Sigma_u^+ \rightarrow X^2\Pi_g^+ \ (0\rightarrow0)$ | 391.4 | 18.7 | [9] |
| | $B^2\Sigma_u^+ \rightarrow X^2\Pi_g^+ \ (1\rightarrow2)$ | 423.65 | | [10] |
| | $B^2\Sigma_u^+ \rightarrow X^2\Pi_g^+ \ (0\rightarrow1)$ | 427.81 | | [10] |
| $N_2$ | $C^3\Pi_u \rightarrow B^3\Pi_g \ (0\rightarrow0)$ | 337.1 | 11.1 | [9] |
| | $C^3\Pi_u \rightarrow B^3\Pi_g \ (1\rightarrow2)$ | 353.67 | | [8] |
| | $C^3\Pi_u \rightarrow B^3\Pi_g \ (0\rightarrow1)$ | 357.69 | | [10] |
| $N_2^*$ | | 337.1 | | [11] |
| | | 357.6 | | [12] |
| N-1 | $2s^22p^2(^3P)3p \rightarrow 2s^22p^2(^1D)5d$ | 425.01 | | [8] |
| NH | $A^3\Pi \rightarrow X^3\Sigma^- \ (0\rightarrow0)$ | 336.0 | 3.7 | [9] |

## The Powder X-Ray Diffaction and Structural Analysis

The powdered diffractogram of $\alpha'$-Fe and $\varepsilon$-Fe$_3$N were shown in Figure 3 and 4. The black signs (+) represent the original diffractogram, the red full line is the analysis result according to *FullProf Suite* program [13] of *Rietveld* [14,15] analysis (30-90 degree of 2 theta) and the blue bar-lines represents the expected space group of the complex. The blue curve indicates the different between the original diffractogram and the analysis result. As shown in Fig. 3 and 4, it is clear that the red full line does almost pass through the black signs (+) original data, and it is demonstrated by the almost flat blue curve indicating that the corresponding analysis is almost perfect. The results of lattice parameters of the crystal structure detailed are presented in Table 3-5. After [1], no nitrides were synthesized in the lower temperature plasma nitriding; the nitrogen atoms were present as a solute in the Fe/Cr lattices by occupying their vacancy sites.

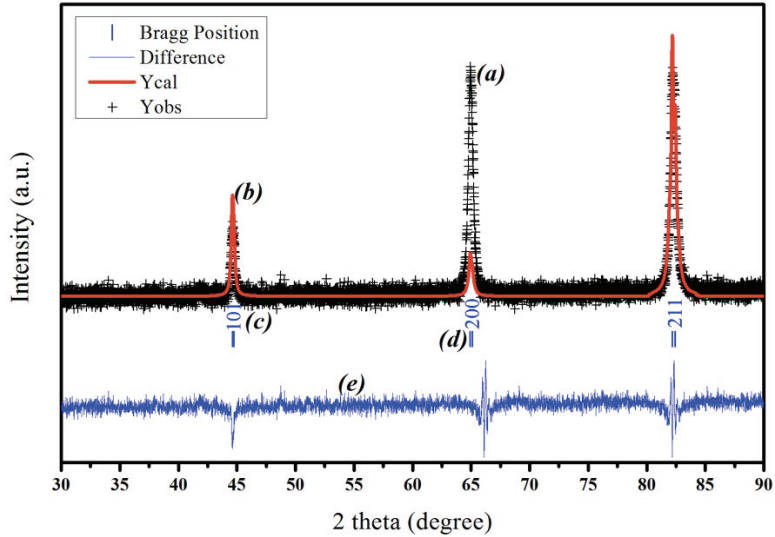

**FIGURE 3.** Diffractogram of $\alpha'$-Fe (black sign +, *a*), and of cubic space group of Im-3m (229) (red full line, *b*), with it's Bragg posision of 2 theta (blue bars, *d*), diffraction lines (blue text of Miller indices hkl, *c*), and and the difference of intensity between the black sign + and the red full line (blue curve, *e*)

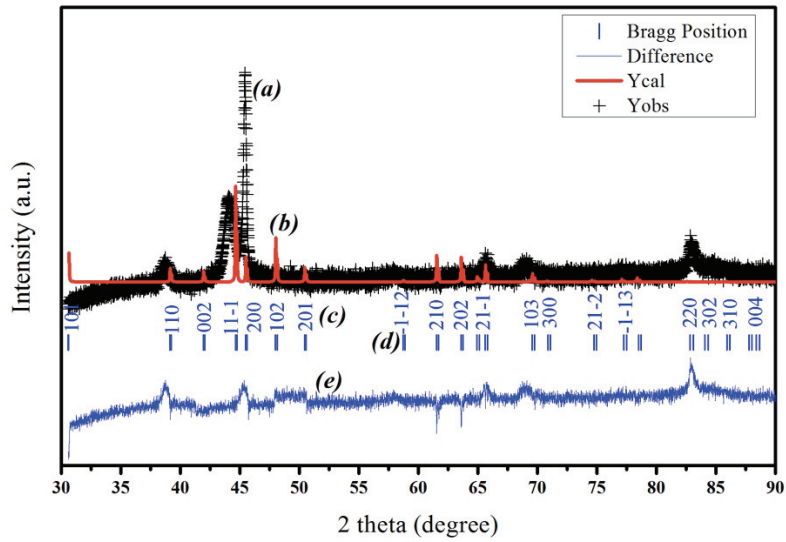

**FIGURE 4**. Diffractogram of ε-Fe₃N (black sign +, *a*), and of trigonal-hexagonal axes space group of P312 (149) (red full line, *b*), with it's Bragg posision of 2 theta (blue bars, *d*), diffraction lines (blue text of Miller indices hkl, *c*), and and the difference of intensity between the black sign + and the red full line (blue curve, *e*)

**TABLE 3**. Detailed lattice parameters of *α'*-Fe and *ε*-Fe₃N due to *FullProf Suite* program [13] of *Rietveld* [14,15] analysis

| Parameter | *α'*-Fe [16] | *ε*-Fe₃N [16] |
|---|---|---|
| Crystal system | cubic | trigonal-hexagonal axes |
| Space group | Im-3m (229) | P312 (149) |
| I/Icor | 13.270000 | 8.310000 |
| *a* (Å) | 2.8680 | 4.6680 |
| *b* (Å) | 0.0000 | 0.0000 |
| *c* (Å) | 0.0000 | 4.3620 |
| *α* (°) | 90.0000 | 90.0000 |
| *β* (°) | 90.0000 | 90.0000 |
| *γ* (°) | 90.0000 | 120.0000 |
| Z | 2.0000 | 2.0000 |
| Calc. density (g/cm³) | 7.86200 | 7.32300 |

**TABLE 4**. Detailed atom coordinates of *α'*-Fe and *ε*-Fe₃N due to crystallography open database

| Phase | Element | Oxid. | *x* | *y* | *z* | Bi | Focc |
|---|---|---|---|---|---|---|---|
| *α'*-Fe [16] | Fe | | 0.000 | 0.000 | 0.000 | 1.000000 | 1.000000 |
| *ε*-Fe₃N [16] | Fe | 1.0 | 0.000 | 0.333 | 0.250 | 1.000000 | 1.000000 |
| | N | -3.0 | 0.000 | 0.000 | 0.000 | 1.000000 | 1.000000 |
| | N | -3.0 | 0.333 | 0.667 | 0.500 | 1.000000 | 1.000000 |

**TABLE 5**. Detailed diffraction lines Miller indices hkl of *α'*-Fe and *ε*-Fe₃N diffractogram

| Phase | d (Å) | h | k | l |
|---|---|---|---|---|
| *α'*-Fe [16] | **2.0280** | **1** | **0** | **1** |
| | 1.4340 | 2 | 0 | 0 |
| | 1.1709 | 2 | 1 | 1 |
| *ε*-Fe₃N [16] | 2.3340 | 1 | 1 | 0 |
| | **2.0579** | **1** | **1** | **-1** |
| | 2.0213 | 2 | 0 | 0 |
| | 1.9195 | 1 | 0 | 2 |

XRD analysis was performed to investigate whether the iron and iron nitrides were precipitated by plasma nitriding at 673 K for 5.4 ks for AISI420 sheet-sample. Besides for the original $\alpha$'-Fe peak at $2\theta = 44.7^o$ (cubic space group), a new peak was detected around $40^o < 2\theta < 45^o$ at $2\theta = 43.52^o$ (trigonal-hexagonal axes space group). This peak is related to the extended martensitic lattice by high nitrogen extraordinary solid solution just as stated in [1] and [17]. Structure of the $\alpha$'-Fe and $\varepsilon$-Fe$_3$N was shown in Figure 5.

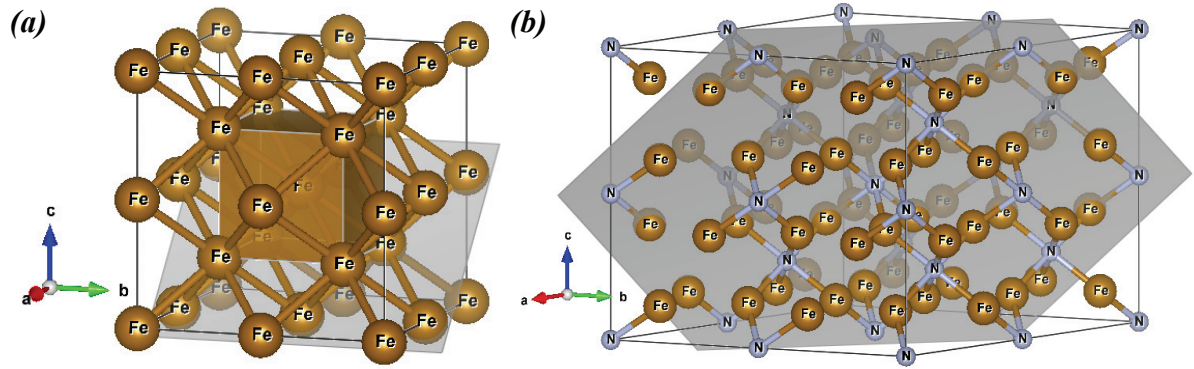

**FIGURE 5.** Schematic picture of the structure of the $\alpha$'-Fe (cubic, *a*) and $\varepsilon$-Fe$_3$N (trigonal-hexagonal axes, *b*) obtained using *Visualization for Electronic and STructural Analysis* (*VESTA*) v.3.3.2 64 bit [18]

## Selective Nitriding and Hardening

Geometry and dimensions are observed and measured by SEM after ink jet printing and plasma nitriding in the present plasma N$_2$-H$_2$. Energy Dispersive X-ray analysis was utilized to describe the nitrogen content distribution at the vicinity of the border between the unprinted surface. The unprinted surface areas in the nitrided micro-texture were selectively nitrided to have high nitrogen solute contents up to 12 mass% in the case of the positive-image and 10 mass% contents in the case of negative image. As depicted in Figure 6 and 7, the nitrogen maps selectively on the unprinted surface area both in the positive and negative images. No nitrogen is present on the printed areas. On the other hand, the carbon from the ink maps only on the printed surface area in both images. This exclusive carbon-nitrogen mapping proves that only the unprinted parts of substrate should have high content nitrogen solutes.

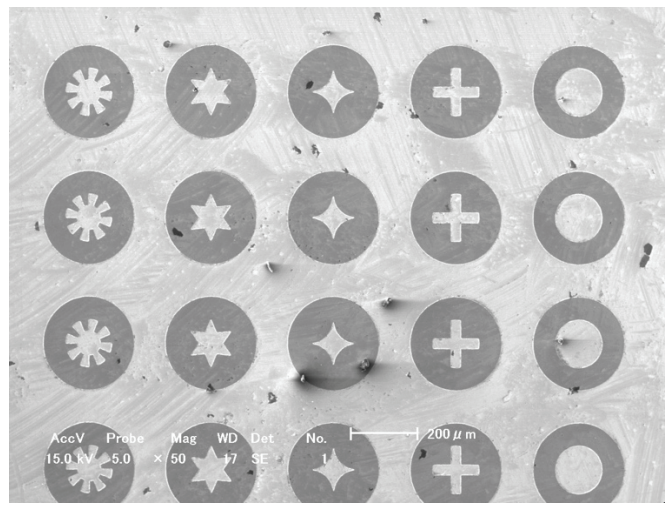

**FIGURE 6.** Printed positive micro-patterned model onto the AISI420

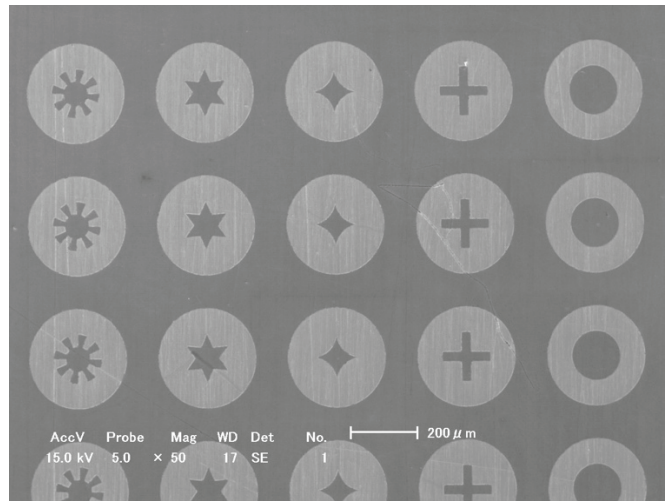

**FIGURE 7**. Printed negative micro-patterned model onto the AISI420

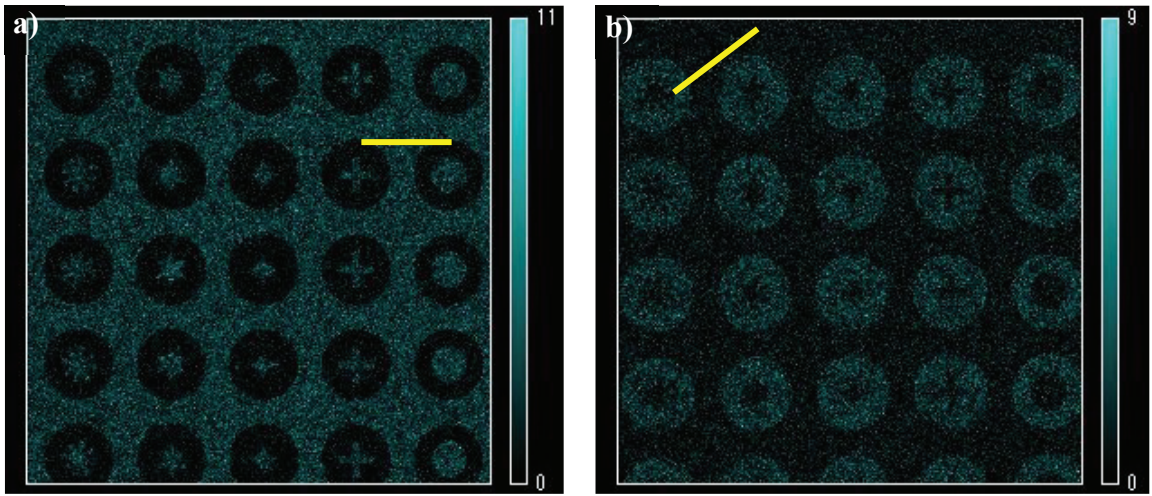

**FIGURE 8**. Nitrogen distribution of micro-patterns: a) positive model onto the AISI420 surface after nitriding; b) negative model onto the AISI420 surface after nitriding

The hardness profile was measured along a single yellow scanning horizontal line across the pattern between the printed and unprinted regions. In order to demonstrate that the original printed pattern should be homogeneously transformed into the hardness profile pattern, a single square region was only left unprinted at the center of AISI420 specimen as shown in Figure 8a and 8b.

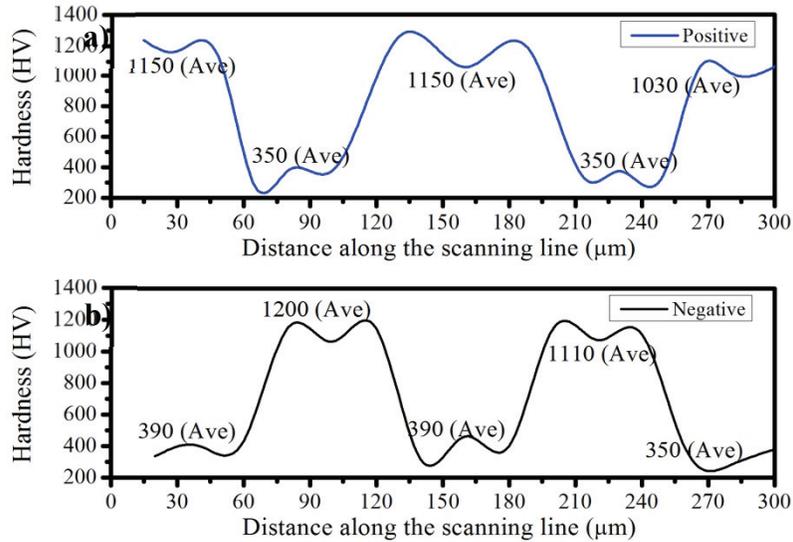

**FIGURE 9**. Transformation of two dimensional micro-patterns to the hardness profile by the present plasma printing for 5.4 ks at 673 K: a) measured hardness profile along the yellow scanning horizontal line in Figure 8a; b) measured hardness profile along the yellow scanning diagonal line in Figure 8b

Figure 9a and 9b compares the hardness profiles measured both in the lateral and longitudinal directions across the mask. Less significant difference was seen in both hardness profiles; i.e. the hardness in the un-printed regions is 1200 Hv, and, it remains to be 350 Hv in the printed region. The steep change of hardness across the edge of masks reveals that the unprinted regions are nitrided to have much higher hardness than matrix hardness of AISI420 stainless steels. The printed regions are free from infiltration of nitrogen atoms into matrix.

## CONCLUSION

The selective nitriding and hardening consists of the initial two dimensional micro-patterning onto the stainless steel specimen and molds and the high density of RF-DC Plasma $N_2$-$H_2$. Through optimization of the nano-carbon ink contents, the printed patterns have sufficient heat resistivity even during plasma nitriding at at 673 K for 5.4 ks by 70 Pa with use of the hollow cathode device. Besides for the original $\alpha'$-Fe peak at $2\theta = 44.7^{o}$ (cubic space group), a new peak was detected around $40^{o} < 2\theta < 45^{o}$ at $2\theta = 43.52^{o}$ (trigonal-hexagonal axes space group). This peak is related to the extended martensitic lattice by high nitrogen extraordinary solid solution. The unprinted surface areas were selectively nitrided to have high nitrogen solute contents up to 12 mass%. The masked or printed area, which has much lower hardness than the nitrided area. The hardness on the unprinted surface was 1200 Hv while it remained to be 350 Hv on the printed surface. The selective nitriding enabled us to construct the three dimensionally textured miniature dies and products by chemical etching of the printed area.

## ACKNOWLEDGEMENTS

The authors would like to express their gratitude to Prof. T. Aizawa, Dr. E.E. Yunata, Mr. A. Farghari, and Mr. S. Kurozumi from Shibaura Institute of Technology (Japan), Dr. K. Wasa from TECDIA, Co. Ltd. (Japan), and Mr. H. Morita from R & D Lab: Thin Film and Nano Coating Technology (Japan) for their help in experiments. The authors participated in this conference was financially supported by the LPDP (Indonesia Endowment Fund for Education), Ministry of Finance, Indonesia.